\documentclass[reprint,showkeys,showpacs,superscriptaddress,nofootinbib,longbibliography,floatfix]{revtex4-2}
\usepackage[utf8]{inputenc}
\usepackage{textgreek}
\usepackage{comment}
\usepackage{lipsum}
\usepackage{natbib}
\usepackage{amsmath}
\usepackage{longtable}
\usepackage{orcidlink}
\usepackage{rotating}
\usepackage{soul}
\usepackage{amssymb}
\usepackage{newunicodechar}
\newunicodechar{∼}{\sim}
\usepackage{enumerate}
\usepackage{array}
\usepackage{newunicodechar}
\newunicodechar{₃}{$_3$}
\usepackage{newunicodechar}
\newunicodechar{₂}{$_2$}
\newunicodechar{−}{-}
\usepackage{color}
\usepackage{tabu}
\usepackage{booktabs}
\usepackage{multirow}
\usepackage{graphicx}
\usepackage{simplewick}
\usepackage[dvipsnames]{xcolor} 
\usepackage[margin=1in]{geometry}
\usepackage{bigints}
\usepackage{dirtytalk}
\usepackage{hyperref}
\definecolor{dark-red}{rgb}{0.9,0.15,0.15}
\definecolor{dark-blue}{rgb}{0.15,0.15,0.4}
\definecolor{medium-blue}{rgb}{0,0,0.5}
\hypersetup{
	colorlinks=true,
	citecolor=blue,
	linkcolor=red,
	filecolor=blue,      
	urlcolor=blue,
}
\begin{document}

\title{Probing Interfacial Magnetic Anisotropy in \texorpdfstring{CoV$_{2}$O$_{4}$}{CoV2O4} using Spin Hall Magnetoresistance}

\author{Sairam Ithineni\,\orcidlink{0009-0006-6986-7317}}
\email{sairam.i@iopb.res.in}
\affiliation{Institute of Physics, Sachivalaya Marg, Bhubaneswar-751005, India}
\affiliation{Homi Bhabha National Institute, Training School Complex, Anushakti Nagar, Mumbai 400094, India}

\author{Krishna Jha\,}
\affiliation{Department of Physics, Indian Institute of Science Bangalore-560012, India}

\author{Aditya A. Wagh\,}
\affiliation{Department of Physics, Indian Institute of Science Bangalore-560012, India}

\author{Shwetha G. Bhat\,\orcidlink{0000-0003-4237-9082}}
\affiliation{Department of Physics, Indian Institute of Science Bangalore-560012, India}

\author{Debashree Nayak}
\affiliation{Homi Bhabha National Institute, Training School Complex, Anushakti Nagar, Mumbai 400094, India}
\affiliation{School of Physical Sciences, National Institute of Science Education and Research, Jatni 752050, India}

\author{K. Senapati\,\orcidlink{0000-0003-1422-3195}}
\affiliation{Homi Bhabha National Institute, Training School Complex, Anushakti Nagar, Mumbai 400094, India}
\affiliation{School of Physical Sciences, National Institute of Science Education and Research, Jatni 752050, India}

\author{P.S. Anil Kumar\,\orcidlink{0000-0002-4574-0868}}
\affiliation{Department of Physics, Indian Institute of Science Bangalore-560012, India}

\author{D. Samal\,\orcidlink{0000-0003-2618-4445}}
\email{dsamal@iopb.res.in}
\affiliation{Institute of Physics, Sachivalaya Marg, Bhubaneswar-751005, India}
\affiliation{Homi Bhabha National Institute, Training School Complex, Anushakti Nagar, Mumbai 400094, India}

\date{\today}

\begin{abstract}
Spin Hall magnetoresistance (SMR) has emerged as a powerful probe for investigating interfacial spin transport and magnetic anisotropy in complex oxide heterostructures. In this work, we investigate the interfacial magnetic anisotropy in Pt/CVO through angle-dependent magnetotransport measurements. Unlike the bulk-sensitive magnetic measurements on both strained CVO and Pt/CVO films, which exhibit a ferrimagnetic transition at $T_{C} \approx 150$ K accompanied by out-of-plane anisotropy that reorients toward in-plane anisotropy below 90 K, SMR reveals a distinct interfacial magnetic anisotropy. The rotational scans of the in-plane transverse SMR at 20 K exhibit substantial hysteresis about [100], while no hysteresis is observed along [110] and [1$\bar{1}$0], indicating a biaxial anisotropy with easy axes along [110] and [1$\bar{1}$0]. Furthermore, the absence of sharp discontinuities in both the in-plane longitudinal and transverse SMR, together with pronounced discontinuities near the in-plane [010] direction during out-of-plane rotation, strongly indicates the presence of in-plane anisotropy. This behavior persists up to 120 K. The discrepancy between the bulk-sensitive magnetic measurements and the SMR response suggests that the Pt/CVO interface retains a magnetic anisotropy distinct from the bulk, highlighting the interfacial sensitivity of SMR. Additionally, the spin mixing conductance is found to be of the order of $10^{14}$ $\Omega^{-1}\mathrm{m}^{-2}$, comparable to other oxide-based spintronic systems. These findings highlight the crucial role of interfacial effects in spin transport and establish Pt/CVO as a promising platform for spintronic applications.
\end{abstract}
\maketitle

\section{Introduction}
Spin Hall magnetoresistance (SMR) has emerged as a powerful electrical probe for investigating \textcolor{black}{complex spin structures} and magnetic anisotropy configurations across a broad range of magnetic systems, including collinear\cite{ji2017spin,lin2017electrical,hou2017tunable,fischer2020large,ishikawa2023spin} and noncollinear antiferromagnets\cite{oda2019magnetoresistance,uchimura2025unconventional}, canted ferrimagnets\cite{ganzhorn2016spin,dong2018spin}. The SMR effect originates from the modulation of the resistivity of a heavy metal layer through spin charge interconversion processes when it is interfaced with an adjacent magnetic material. \textcolor{black}{In heavy metals with strong spin-orbit coupling, such as Pt, Ta, and W, an applied charge current generates a transverse spin accumulation through the spin Hall effect, enabling efficient spin-current generation in spintronic heterostructures\cite{vila2007evolution,liu2012spin,hao2015giant}. Owing to its interface-sensitive nature, SMR plays a pivotal role in the Rashba Edelstein effect\cite{nakayama2016rashba}, the spin-momentum locking of topological surface states\cite{chen2022control}, the orbital Hall and Rashba-Edelstein effects\cite{ding2020harnessing}, while serving as a complementary approach to sophisticated techniques such as synchrotron radiation and neutron diffraction\cite{chen2016theory}.} Furthermore, SMR enables extraction of the interfacial spin-mixing conductance, a key parameter governing spin transport efficiency and spin-current transfer across interfaces\cite{onbasli2014pulsed,qiu2013spin,wang2014scaling}. \textcolor{black}{SMR} has been extensively investigated in heavy metal/magnetic insulator bilayers, particularly Pt/Y$_3$Fe$_5$O$_{12}$ (YIG), owing to the exceptionally low Gilbert damping parameter of YIG. Although garnets have been widely explored for spin transport studies, their integration with existing microelectronics remains challenging. In contrast, complex oxide spinel magnetic insulators such as Ni$_{0.65}$Zn$_{0.35}$Al$_{0.8}$Fe$_{1.2}$O$_4$ (NZAFO) and MgAl${0.5}$Fe$_{1.5}$O$_4$ (MAFO)\cite{emori2017coexistence,emori2018ultralow,gray2018spin} are compatible with silicon-based technologies\cite{kang2006epitaxial,warusawithana2009ferroelectric} and have shown efficient spin pumping and electrical spin-current detection in bilayers\cite{gray2018spin}, making spinel oxides attractive for spintronic devices.

\textcolor{black}{In the family of spinel oxides}, vanadates with the general formula AV$_2$O$_4$ constitute a compelling class of materials in which the \textcolor{black}{interplay of} strong electronic correlations, geometric frustration, orbital ordering, gives rise to a wide variety of emergent phenomena\cite{kawaguchi2016orthorhombic,niitaka2013type,suzuki2007orbital}. CoV$_2$O$_4$ (CVO) stands out among spinel vanadates due to its proximity to itinerant behavior, arising from the short V–V distance near the critical limit for metallicity\cite{kismarahardja2011co,kiswandhi2011chemical}. \textcolor{black}{Bulk} CVO crystallises in a cubic spinel structure, where Co$^{2+}$ ($3d^7$) ions occupy the tetrahedral A-sites, while V$^{3+}$ ($3d^2$) ions reside at the octahedral B-sites. Magnetically, cubic bulk CoV$_2$O$_4$ (CVO) exhibits a nearly collinear ferrimagnetic (CF) ground state below $T_{\mathrm{C}} \sim 150$ K and undergoes a subtle first-order structural transition near $\sim 90$ K, accompanied by the emergence of spin canting\cite{reig2016structural} whereas CVO thin films exhibit pronounced strain-dependent magnetic anisotropy and spin reorientation behavior.
 Under compressive strain [SrTiO$_3$ (STO) and MgAl$_2$O$_4$ (MAO)], the magnetization \textcolor{black}{below $\sim 150$ K ($\sim 160$ K for MAO)}  favours along the out-of-plane direction below $\sim 150$ K ($\sim 160$ K for MAO) and subsequently reorients toward the in-plane direction near $\sim 90$ K ($\sim 160$ K for MAO)\cite{behera_2021_cvo,khabchi_2026_cvo,liu_2024_cvo}. While under the tensile strain (MgO), magnetization switches from an in-plane at $\sim 127$ K to an out-of-plane easy axis below $\sim 45$ K\cite{behera_2021_cvo,khabchi_2026_cvo,liu_2024_cvo}.
 \textcolor{black}{Further}, CVO grown on SrTiO$_3$ exhibits a field-driven spin reorientation below $\sim 75$ K, where the easy axis changes from [110] at low fields to [100] above $\sim 3$ T, accompanied by V-site spin canting\cite{thompson_2020_thesis}. The temperature and field-dependent evolution of magnetic anisotropy in CVO/STO is summarised in the phase diagram shown in Fig.~1. Complementarily, Kim et al. investigated CVO/STO using torque magnetometry to probe its magnetic anisotropy. Interestingly, while earlier zero-field neutron measurements identified the low-temperature easy axis along [110], torque magnetometry under applied magnetic fields revealed that the in-plane $\langle100\rangle$ directions are energetically favourable\cite{kim_2023_cvo}. These contrasting observations point toward a complex evolution of magnetic anisotropy in CVO. Such tunable magnetic anisotropy, together with the temperature-driven spin reorientation in CVO, makes it an attractive platform for spintronic applications, as it can facilitate efficient manipulation of spin configurations. \par
 \begin{figure*}
\centering
\includegraphics[width=\linewidth]{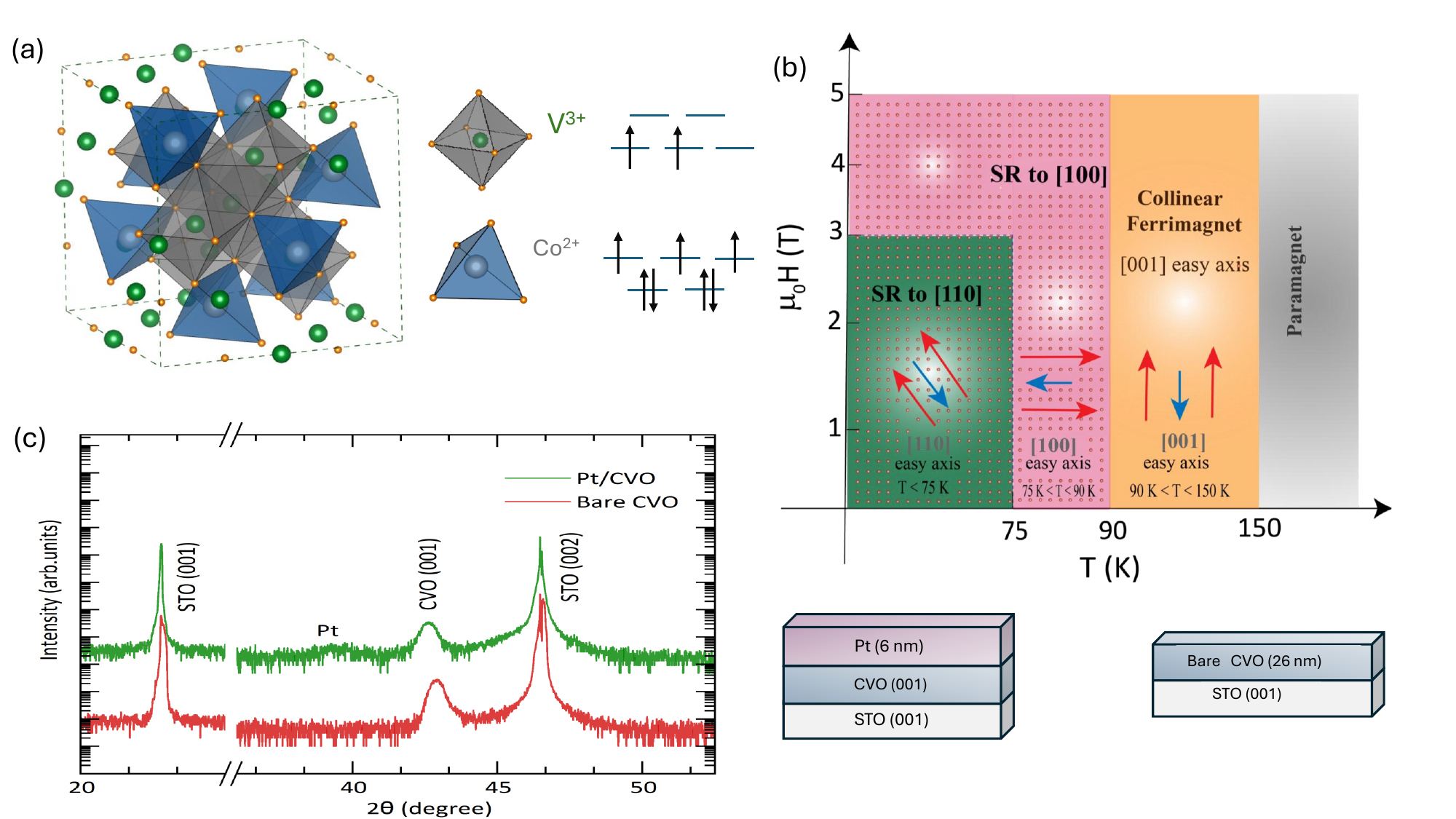}
\caption{\textcolor{black}{(a) Crystal structure of CoV$_2$O$_4$ (CVO), illustrating the arrangement of CoO$_4$ tetrahedra and VO$_6$ octahedra along with their corresponding electronic configurations and spin states of Co$^{2+}$ and V$^{3+}$ ions. (b) Schematic magnetic phase diagram of CVO, illustrating the evolution of magnetic anisotropy and spin configurations with temperature. The red and blue arrows represent the spin orientations of Co and V ions, respectively, while the dotted arrows indicate the canting of V spins in the spin-reoriented phase. (c) $\theta$--$2\theta$ X-ray diffraction patterns of the Pt/CVO bilayer and bare CVO thin films grown on STO(001), showing the CVO (00$l$) reflections and the additional Pt diffraction peak from the overlayer. Insets show the Pt (6 nm)/CVO (26 nm) and bare CVO (26 nm) thin-film heterostructures grown on SrTiO$_3$ (001) substrates.}}
 \label{fig:1}
\end{figure*}
Although the temperature and field dependent magnetic anisotropy of bulk CVO has been extensively studied, its interfacial magnetic behaviour remains largely unexplored. To gain insight into interfacial magnetic anisotropy and spin transport in CVO, SMR measurements were carried out on Pt/CVO heterostructures. Prior to the SMR studies, magnetic measurements performed on bare CVO/STO and Pt/CVO/STO samples revealed similar anisotropic behavior, suggesting that Pt deposition does not significantly affect the intrinsic anisotropy of CVO. Subsequently, SMR measurements were  performed under both in-plane and out-of-plane magnetic field configurations to investigate interfacial magnetic behavior and spin transport. The low-temperature in-plane SMR response at 20 K reveals the presence of biaxial anisotropy with easy axes along the [110] and [1$\bar{1}$0] directions. Furthermore, the SMR results confirm the persistence of in-plane anisotropy up to 110 K, unlike the magnetic measurements on Pt/CVO, likely due to the presence of distinct surface anisotropy in CVO. Given the interface-sensitive nature of SMR, the measured response likely reflects the surface magnetic behavior rather than the bulk. Moreover, the spin mixing conductance parameter estimated from the SMR measurements, is on the order of $2 \times 10^{14}~\Omega^{-1},\mathrm{m}^{-2}$, indicating that the Pt/CVO interface is highly transparent to spin currents. Overall, these results demonstrate that SMR provides valuable insight into the interfacial magnetic behavior and spin transport properties of CVO heterostructures.

\section {Experimental Methods :}

 CVO thin films were grown on STO(001) substrates using PLD, with a KrF excimer laser (λ = 248 nm), operating at a repetition rate of 4 Hz and a laser fluence of 1.8 J/cm$^2$. A CoV$_{2}$O$_{6}$ target was used, and the substrate temperature was maintained at 650$^0$ C under a base pressure of 2$\times$ 10$^{−6}$ mbar. Subsequently, a Pt layer was deposited on top of the CoV$_2$O$_4$ layer  by sputtering at room tempearture to avoid interdiffusion. The structural characterization of the films was carried out using a Rigaku SmartLab high-resolution X-ray diffractometer (HR-XRD) with a Cu K$\alpha_1$ radiation source ($\lambda = 1.54~\text{\AA}$). $\theta$–$2\theta$ scans were performed to confirm the phase, while X-ray reflectivity (XRR) measurements were carried out to determine the film thickness. DC Magnetization measurements were carried out using a \textcolor{black}{superconducting quantum interference device-based magnetometer (Quantum Design SQUID-VSM)}. The films were then patterned into Hall bars using UV photolithography. Electrical contacts were made by using wire-bonded with Au|cr wires. For SMR measurements, The sample was mounted in a modified closed-cycle refrigerator equipped with a rotating magnet arrangement. The angular dependence of SMR was measured by rotating the sample in both clockwise and counterclockwise directions at various constant magnetic fields in the range of 150 Oe to 3.5KOe. An AC current ($I_{\mathrm{P-P}} = 2$ mA, 333 Hz) was applied using a Keithley 6221 DC/AC current source, and the transverse voltage was measured using a Stanford Research SR830 lock-in amplifier. 
 
\begin{figure*}
\centering
\includegraphics[width=\linewidth]{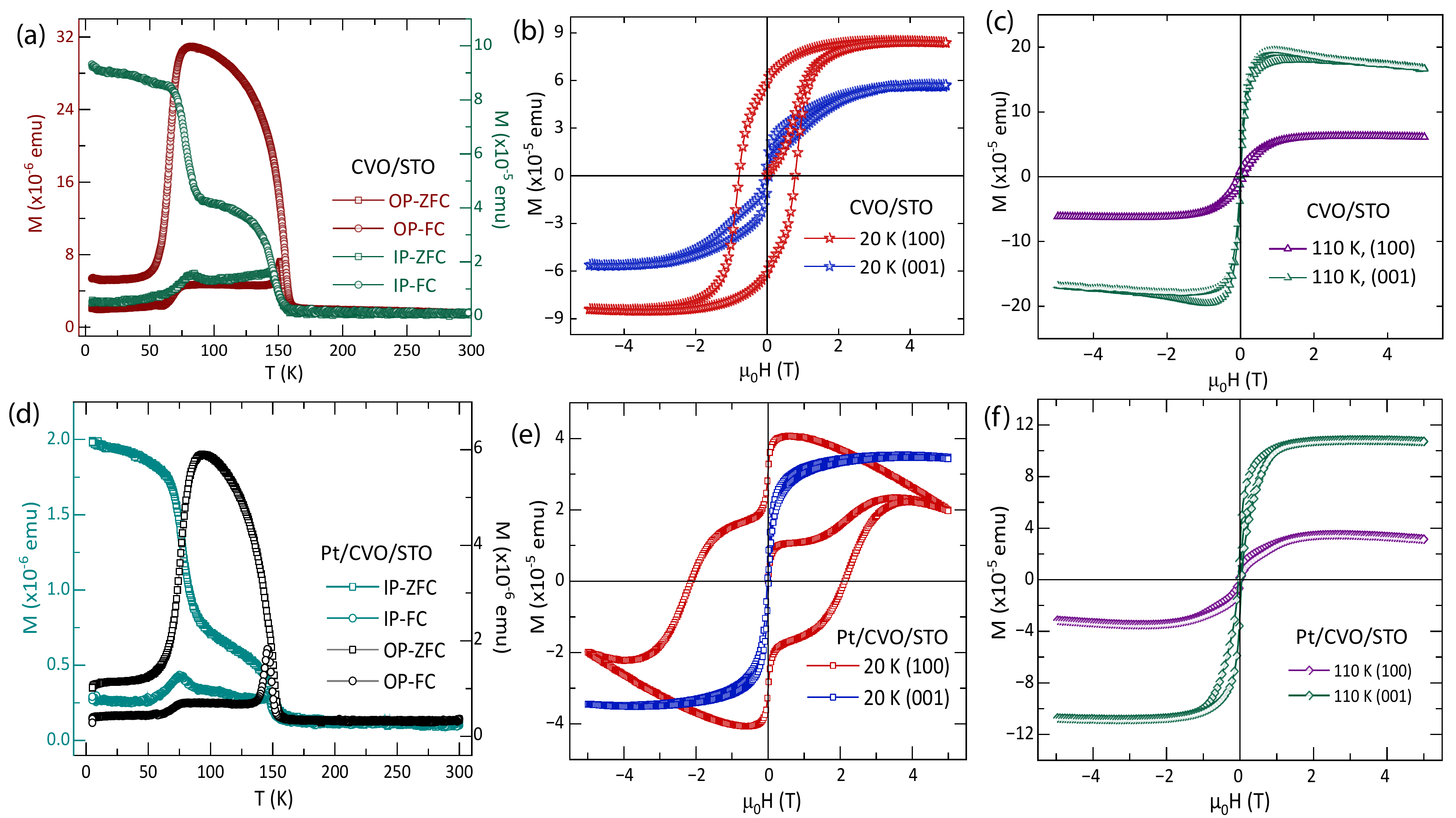}
\caption{\textcolor{black}{Temperature- and field-dependent magnetic properties of bare CVO and Pt/CVO heterostructures. The upper panel [(a)--(c)] corresponds to bare CVO, while the lower panel [(d)--(f)] represents Pt/CVO. Panels (a) and (d) show the temperature dependence of the magnetization measured under zero-field-cooled (ZFC), field-cooled cooling (FCC), and field-cooled warming (FCW) protocols. Panels (b), (c), (e), and (f) display the magnetic field dependence of magnetization, $M(H)$, measured at different temperatures, highlighting the evolution of magnetic hysteresis and saturation behavior in both systems.}}
 \label{fig:1}
\end{figure*}
\section{Results and Discussion:} 
Figure 1(b) shows the $\theta$–$2\theta$ scan of representative bare CoV$_2$O$_4$ (26 nm) and Pt (6 nm)/CoV$_2$O$_4$ (26 nm) thin films. \textcolor{black}{The (001) diffraction peaks corresponding to the CVO layer are clearly observed in both samples, indicating oriented growth, while the Pt/CVO sample exhibits an additional peak arising from the Pt overlayer. No impurity phases are detected, confirming that the crystallographic structure of CVO remains preserved after Pt deposition. }\par
Figure 2(a) shows the temperature-dependent magnetisation measured under zero-field-cooled (ZFC) and field-cooled (FC) conditions in an applied field of 100 Oe over the temperature range 5 to 300 K, with the magnetic field applied along the in-plane (IP) and out-of-plane (OP) directions of CVO/STO (001). Consistent with previous reports, the system undergoes a paramagnetic-to-ferrimagnetic transition  near $\sim150$ K ($T_c$)\cite{behera_2021_cvo,thompson2018spin}. A spin reorientation from the out-of-plane (OP) to the in-plane (IP) direction occurs near $\sim90$ K ($T_1$), accompanied by a decrease in OP magnetization and a corresponding increase in IP magnetization.  Figure 2(b) and 2(c) present the field-dependent magnetization measured in both IP and OOP configurations at 20 K and 110 K, respectively, to probe the anisotropy across different magnetic phases. At $T = 20~K$, the sample exhibits a larger coercive field along \textcolor{black}{inplane} compared to \textcolor{black}{out of plane ([001])}, indicating a stronger anisotropy along the inplane direction. In contrast within the ferrimagnetic phase at 110 K, the coercive field becomes nearly negligible, while the comparatively lower field  required to saturate the magnetization along [001] indicates that the [001] direction is the easy axis compared to [100] at 110 K. The temperature-dependent evolution of the magnetic easy axis from \textcolor{black}{out of plane along [001]} at high temperatures to \textcolor{black}{inplane} at low temperatures, as confirmed by the $M$–$H$ measurements, is indicative of a spin reorientation transition. These observations are consistent with evolution of magnetic anisotropy of CVO/STO, as reported earlier\cite{thompson_2020_thesis}. 
To further examine whether the Pt overlayer induces any \textcolor{black}{substantial} modification in the magnetic response, magnetization measurements were performed on Pt/CVO/STO, as shown in Fig.~2(d). The magnetization exhibits anisotropic behavior similar to that of bare CVO/STO, with all characteristic magnetic transitions occurring at nearly identical temperatures. Although the $M$–$H$ curves at 20 K and 110 K exhibit some variation in the coercive field values following Pt deposition, the overall anisotropic characteristics and the temperature-driven spin reorientation behavior remain essentially unchanged in Pt/CVO/STO. These results demonstrate that the deposition of Pt overlayer does not significantly affect the overall magnetic anisotropy of CVO, as confirmed by our magnetization measurements. However, previous neutron studies have reported that below $\sim75$ K, CoV$_2$O$_4$ undergoes spin canting accompanied by a reorientation of spins toward the [110] direction\cite{thompson_2020_thesis}. In the present case, the magnetization measurements are limited to the [100] and [001] orientations, the anisotropy between the [110] and [100] directions could not be conclusively resolved, thereby necessitating complementary angle-sensitive probes \textcolor{black}{such as SMR}, capable of detecting in-plane anisotropic variations.
\begin{figure*}
    \centering
    \includegraphics[width=\linewidth]{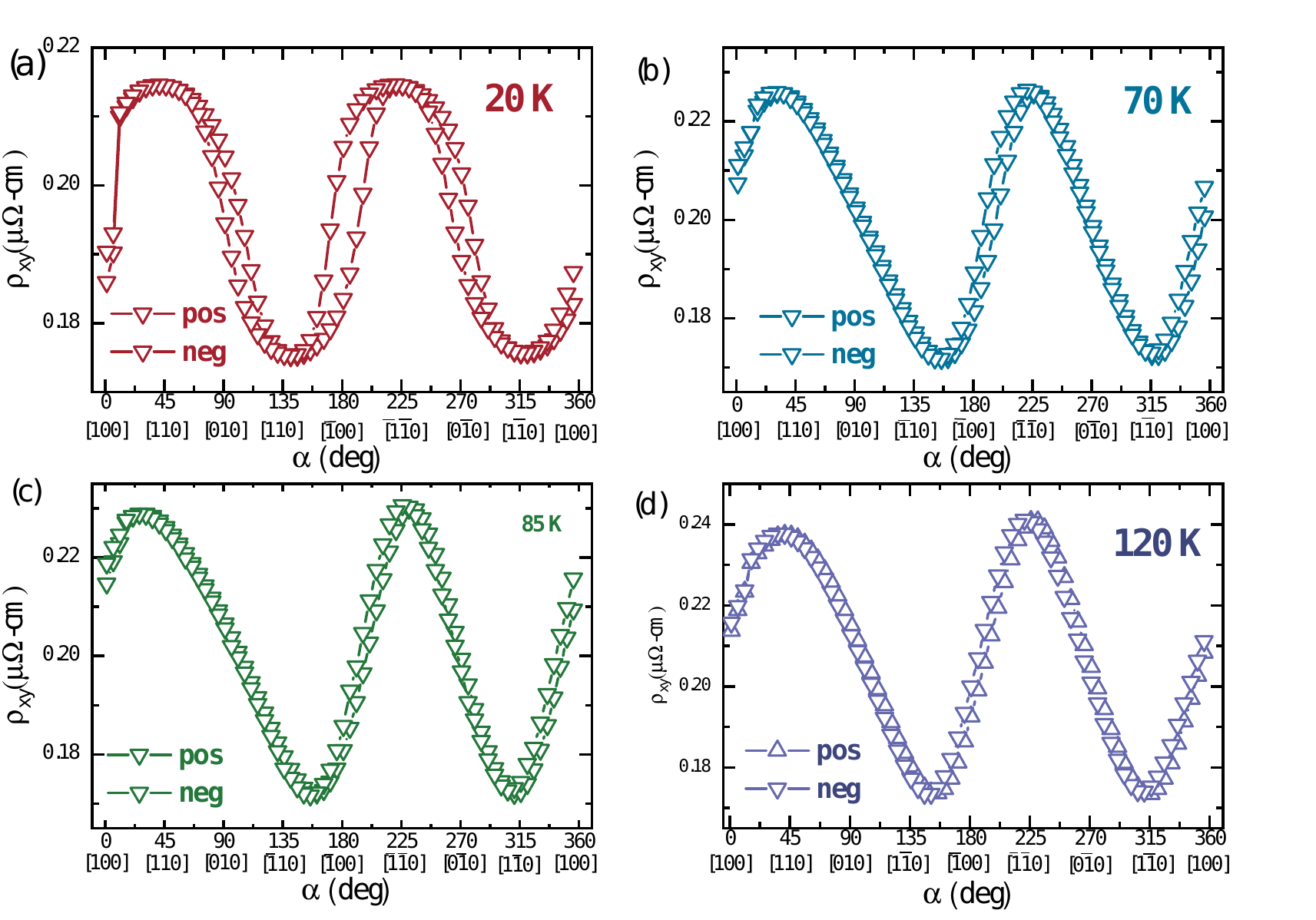}
   \caption{\textcolor{black}{Angular dependence of the transverse resistivity, $\rho_{xy}$, measured under positive and negative magnetic field sweeps at different temperatures for the Pt/CVO heterostructure under an applied magnetic field of 150~Oe. Panels (a)--(d) correspond to measurements performed at 20~K, 70~K, 85~K, and 120~K, respectively. The sinusoidal variation in $\rho_{xy}$ with rotation angle $\alpha$ reflects the anisotropic magnetotransport response and its evolution with temperature.}}
    \label{fig:1}
\end{figure*}

\textcolor{black}{Therefore to illucidate the underlying anisotropy,  SMR measurements were carried out on Pt/CVO  using angle-dependent magnetoresistance (ADMR).} In the Pt/CoV$_2$O$_4$ device, when a charge current is applied along the $x$-direction in the Pt Hall bar, the spin Hall effect (SHE) generates a transverse spin current, leading to spin accumulation ($\mu_s$) at the Pt/CoV$_2$O$_4$ interface. The relative orientation between the spin polarization ($\mathbf{s}$) in Pt and the magnetization $\mathbf{M}$ in CoV$_2$O$_4$ governs the absorption of the spin current \textcolor{black}{mediated by} spin-transfer torque at the interface. This results in a modulation of the spin current in the Pt layer and, consequently, a change in charge current via the inverse spin Hall effect (ISHE), \textcolor{black}{thereby producing a modulation in the resistance of the Pt layer upon rotation of the applied magnetic field.} The longitudinal resistivity $(\rho_{xx})$ and the transverse resistivity $(\rho_{xy})$ measured in the Pt exhibits an angular dependence as \cite{althammer2013quantitative,chen2013theory,chen2016theory} : 
 \begin{equation}
     \rho_{xx} = \rho_{0}+\Delta \rho_{1}(m^2_{y})
\end{equation}
\begin{equation}
    \rho_{xy} = \rho_{2}m_{z}+\rho_{1}m_{x}m_{y}
\end{equation}
where $m(m_{x},m_{y},m_{z}) = m/m_{S}$ are the direction cosines of magnetizations along x,y,z. Here $\rho_{0}$ is the base line resistivity , $\frac{\rho_{1}}{\rho_{0}}$ is SMR, and $\rho_{2}$ is an anomalous Hall like contribution. \par
\begin{figure*}
    \centering
    \includegraphics[width=\linewidth]{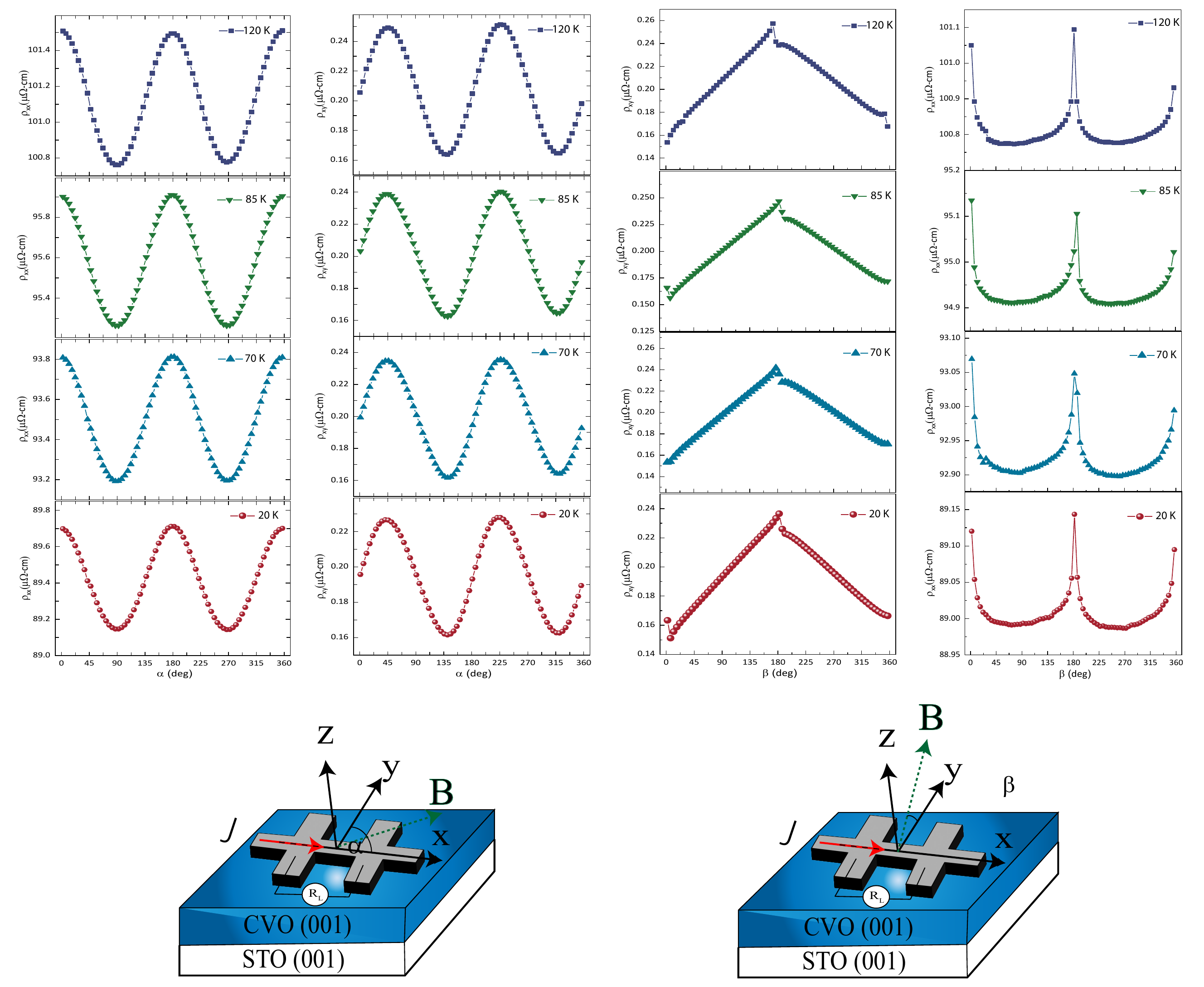}
   \caption{\textcolor{black}{Angular-dependent magnetoresistance measurements of the Pt/CVO heterostructure under different magnetic field rotation geometries at various temperatures. The left panel shows the angular dependence of the longitudinal resistivity, $\rho_{xx}$, measured during in-plane magnetic field rotation, while the right panel presents measurements obtained under out-of-plane field rotation. Data are shown for temperatures ranging from 20~K to 120~K. The corresponding schematics at the bottom illustrate the measurement configurations, including the current direction ($J$), magnetic field orientation ($B$), and crystallographic axes of the Pt/CVO/STO(001) heterostructure. The observed angular modulation in $\rho_{xx}$ demonstrates the anisotropic magnetotransport response and its dependence on temperature and magnetic field orientation.
}}
    \label{fig:1}
\end{figure*}
At a fixed magnetic field $H$, angular-dependent measurements were performed by rotating the field in both forward and reverse directions within the in-plane ($xy$-plane) configuration. The corresponding transverse ($\rho_{xy}$) and longitudinal ($\rho_{xx}$) resistivities were recorded for both $\alpha$ and $-\alpha$ rotation scans. Figure 3 shows the in-plane transverse resistivity ($\rho_{xy}$) as a function of rotation angle $\alpha$ measured at 150 Oe for different temperatures. The magnetic field is initially aligned along the current direction ($\mathbf{B}\parallel \mathbf{j}$ $\parallel$ $x$ axis) and is then rotated continuously within the film plane toward the $y$ axis. The transverse SMR qualitatively follows a $\sin(2\alpha)$ dependence; however, significant deviations from this behavior are observed at all temperatures. Notably, at 20 K, a clear hysteresis between the forward and reverse scans emerges when the applied field is oriented along the [100] direction.\textcolor{black}{The observed hysteresis reflects the presence of hard-axis anisotropy.\cite{wagh2022probing}} \textcolor{black}{As the magnetization approaches along the hard axis direction, it undergoes abrupt switching between orientations on either side of the hard axis}.  On the other hand, the absence of hysteresis corresponds to smooth rotation of magnetization along an easy axis. \textcolor{black}{Thus, the absence of hystersis along the symmetry-equivalent ⟨110⟩ directions, namely [110],and [$\bar{1}\bar{1}$0] together with the orthogonal pair [$\bar{1}$10] and [1$\bar{1}$0], indicate the presence of biaxial in-plane anisotropy at 20 K.} Further, the hysteresis progressively diminishes with increasing the applied field at 20 K, indicating that the Zeeman energy overcomes the magnetic anisotropy, resulting in field-driven magnetization rotation. (See Supplimentary material).\par
\textcolor{black}{Upon increasing the temperature from 20 K to 110 K in the in-plane transverse measurements, the angular dependence progressively deviates from the expected $\sin(2\alpha)$ behaviour, accompanied by a substantial decrease in the hysteresis around the [010] and [0$\bar{1}$0] directions. At these elevated temperatures, the anisotropy along the [100] and [110] directions becomes nearly comparable, making it difficult to clearly distinguish the magnetic easy axis. Furthermore, the absence of sharp discontinuities or abrupt transitions in the rotational transverse measurements indicates the persistence of an in-plane anisotropy. Consistently, no discernible change is observed in the rotational transverse SMR response between 70 K, 85 K, and 110 K, suggesting that the in-plane anisotropic behavior persists up to 110 K.}
 However, the magnetic measurements on Pt/CVO suggest that the system remains in the ferrimagnetic phase with an out-of-plane easy axis at 110~K, while a distinct in-plane anisotropy emerges below 90~K. This discrepancy between the bulk magnetic response and the SMR response could be further examined by performing both transverse and longitudinal SMR measurements in the out-of-plane rotational and in-plane geometries, as shown in Figure~4.\par Figures~4(a) and 4(b) display the angular dependence of the transverse resistivity, $\rho_{xy}$, and longitudinal resistivity, $\rho_{xx}$, respectively, as a function of the rotation angle $\alpha$ under an applied magnetic field of 1000~Oe at different temperatures. Both $\rho_{xy}$ and $\rho_{xx}$ exhibit a clear periodic modulation with a period of $180^\circ$. The longitudinal resistivity shows maxima when the magnetic field is aligned parallel or antiparallel to the current direction ($\alpha = 0^\circ$ and $180^\circ$) and minima when the field is oriented perpendicular to the current direction ($\alpha = 90^\circ$ and $270^\circ$). Such behavior is consistent with the expected $\cos^2\alpha$ dependence of the longitudinal SMR contribution \textcolor{black}{($\rho_{xx} = \rho_0 +\rho_1 cos^2\alpha $)}, indicating that the spin-dependent scattering at the interface is governed by the projection of magnetization relative to the spin polarization direction. Simultaneously, the \textcolor{black}{in-plane} transverse component, $\rho_{xy}^{\mathrm{SMR}}$, exhibits the characteristic $\sin\alpha \cos\alpha$ angular dependence expected for transverse spin Hall magnetoresistance. The extrema in $\rho_{xy}^{\mathrm{SMR}}$ occur near $\alpha = 45^\circ$ and $135^\circ$ for the positive maxima, while minima are observed close to $\alpha = -45^\circ$ and $225^\circ$. \textcolor{black}{Interestingly, the longitudinal and transverse SMR responses remain nearly unchanged under the applied magnetic field across all measured temperatures up to 120~K, suggesting that the in-plane anisotropy observed at 20~K likely persists up to 120~K}. Figure 4 (c) and Figure 4 (d) illustrate the out-of-plane transverse ($\rho_{xy}$) and longitudinal resistivity ($\rho_{xx}$) as a function of rotation angle $\beta$, measured at 1000 Oe for different temperatures. This configuration is chosen such that the signal predominantly contains the SMR contribution, with minimal contribution from anisotropic magnetoresistance (AMR). In this configuration, the magnetic field is initially aligned along the $y$ axis and is rotated toward the $z$ axis, reaching $90^\circ$ at $\mathbf{H} \parallel z$. For both the transverse and longitudinal out of plane resistivity responses, sharp discontinuities are observed when the applied field is \textcolor{black}{approaching} towards the in-plane directions at [010] and [0$\bar{1}$0], indicating the presence of strong in-plane anisotropy. Notably, this behavior is observed across all measured temperatures up to 120 K. \textcolor{black}{Altogether, the absence of sharp discontinuities in the in-plane transverse and longitudinal responses, together with the presence of pronounced discontinuities when approaching the in-plane [010] orientation in the out-of-plane rotational configuration, clearly indicates the existence of in-plane anisotropy. Thus, the ADMR measurements performed in both configurations confirm the presence of a pronounced in-plane magnetic anisotropy that persists up to 120~K.} However, this observation contrasts with our bulk sensitive magnetic measurements, which suggest that the Pt/CVO likely develops an out-of-plane anisotropy at temperatures above 90~K. Since SMR predominantly probes the magnetization at the Pt/CoV$_2$O$_4$ interface, the persistence of in-plane anisotropic features in the SMR response up to higher temperatures suggests that the surface magnetization of CVO may possess a magnetic anisotropy distinct from that of the bulk. Thus, although bulk-sensitive measurements indicate the emergence of out-of-plane anisotropy, the SMR response exhibits a contrasting behaviour dominated by interfacial in-plane anisotropy. Similar interfacial anisotropy-dominated SMR behavior has previously been reported in CoFe$_2$O$_4$-based heterostructures\cite{isasa2016spin}.

\textcolor{black}{Further, we estimated the spin conductance across the Pt/CVO using the eqn.3\cite{chen2013theory}}  
\begin{equation}
\left| \frac{\Delta \rho_{\mathrm{SMR}}}{\rho_0} \right|
=
\frac{
\theta_{\mathrm{SH}}^2
\left(2 \lambda_s \rho_{\mathrm{Pt}}\right)
t_{\mathrm{Pt}}^{-1}
G_r
\tanh^2 \left( \frac{t_{\mathrm{Pt}}}{2\lambda_s} \right)
}{
1 + 2 \lambda_s \rho_{\mathrm{Pt}} G_r
\coth \left( \frac{t_{\mathrm{Pt}}}{\lambda_s} \right)
}.
\end{equation}
In the analysis, the spin Hall angle of Pt, $\theta_{\mathrm{SH}}$, was assumed to lie in the range of $0.07$--$0.124$\cite{weiler2013spin}, while the spin diffusion length, $\lambda_s$, was taken to be $\sim 2.0$--$2.3$~nm based on previous reports on insulating spinel systems within the Dyakonov--Perel (DP) extrinsic scattering framework\cite{riddiford2019efficient,guo2022spin,hui2016spin,pena2023spin}. Using these parameters, the effective spin mixing conductance was determined to be in the range of
\[
G_r \approx (1.2\text{--}3.5)\times10^{14}~\Omega^{-1}\mathrm{m}^{-2}.
\]
The obtained values are comparable to those reported from spin-pumping and damping enhancement measurements in several insulating spinel-based heterostructures\cite{riddiford2019efficient}. In particular, previous studies on MAFO/Pt and MAFO/$\beta$-W bilayers have demonstrated spin mixing conductance values of the order of $\sim 2\times10^{14}~\Omega^{-1}\mathrm{m}^{-2}$, exceeding many of the reported values for YIG/Pt systems. The comparable magnitude of $G_r$ obtained in the present Pt/CVO heterostructure highlights the high transparency of the Pt/CVO interface to spin currents and confirms the suitability of CVO-based heterostructures for efficient spin-current generation, transmission, and manipulation. These results further establish spinel oxide heterostructures as promising candidates for pure spin-current-based spintronic applications employing magnetic insulating platforms. 

\section{Conclusion:}
In summary, spin Hall magnetoresistance (SMR) measurements on Pt/CoV$_2$O$_4$ reveal a robust in-plane magnetic anisotropy that persists across a wide temperature range up to 110 K. Angular-dependent measurements demonstrate clear hysteresis at low fields, identifying an in-plane easy axis along the $\langle110\rangle$ directions, while high-field measurements confirm that the magnetization undergoes field-driven rotation consistent with the SMR framework. Notably, both in-plane and out-of-plane rotation geometries consistently exhibit signatures of in-plane anisotropy even at temperatures where bulk magnetization measurements indicate a transition to out-of-plane anisotropy. This apparent discrepancy highlights the interfacial sensitivity of SMR, suggesting that the Pt/CoV$_2$O$_4$ interface retains a distinct magnetic anisotropy compared to the bulk. The persistence of interfacial in-plane anisotropy, despite bulk reorientation, points to a decoupling between surface and bulk magnetic behavior. These findings underscore the importance of interface-driven effects in spin transport and establish Pt/CoV$_2$O$_4$ as a compelling platform for studying anisotropy engineering and interfacial magnetism in complex oxide heterostructures.

\section*{Acknowledgment}
D.S. acknowledges the funding from Max Planck Partner Group and SERB, Government of India (Grant No. CRG/2019/005144). S.I. thanks Priya Lekshmi, Vikram, and Pranjul for assistance during the measurements. 
\section*{Data Availability Statement}
The data that support the findings of this study are available from the corresponding author upon reasonable request.

\bibliography{CVOrefs.bib}

\end{document}